\documentclass[a4paper,fleqn,usenatbib]{mnras}

\usepackage{amsmath}	% Advanced maths commands
\usepackage{txfonts}
\usepackage{mathrsfs}
\usepackage{breqn}

\usepackage[T1]{fontenc}
\usepackage{ae,aecompl}

\usepackage{graphicx}	% Including figure files
\usepackage{amssymb}	% Extra maths symbols
\usepackage{multirow}
\usepackage{hyperref}
\hypersetup{
    colorlinks,
    citecolor=blue,
    filecolor=black,
    linkcolor=blue,
    urlcolor=black
}

\newcommand{\fesc}{\ifmmode{f_{\rm esc}}\else{$f_{\rm esc}$}\fi}
\newcommand{\fescs}{\ifmmode{f_{\rm esc}^\star}\else{$f_{\rm esc}^\star$}\fi}
\newcommand{\kms}{\ifmmode{{\;\rm km~s^{-1}}}\else{km~s$^{-1}$}\fi}
\newcommand{\fgas}{\ifmmode{{f_{\rm gas}}}\else{$f_{\rm gas}$}\fi}
\newcommand{\cubecm}{\ifmmode{{\rm cm^{-3}}}\else{cm$^{-3}$}\fi}
\newcommand{\ztwo}{\ifmmode{{\rm [Z_2/H]}}\else{[Z$_2$/H]}\fi}
\newcommand{\zthree}{\ifmmode{{\rm [Z_3/H]}}\else{[Z$_3$/H]}\fi}
\newcommand{\lsim}{\lower0.3em\hbox{$\,\buildrel <\over\sim\,$}}
\newcommand{\gsim}{\lower0.3em\hbox{$\,\buildrel >\over\sim\,$}}

\newcommand{\sfr}{\ifmmode{\textrm{M}_\odot \,\textrm{yr}^{-1} \,\textrm{Mpc}^{-3}}\else{M$_\odot$ yr$^{-1}$ Mpc$^{-3}$}\fi}
\newcommand{\hsfr}{\ifmmode{\textrm{M}_\odot\, \textrm{yr}^{-1}}\else{M$_\odot$ yr$^{-1}$}\fi}

\newcommand{\eavg}{\ifmmode{\langle E_\gamma \rangle}\else{$\langle E_\gamma \rangle$}\fi}

\newcommand{\enzo}{{\sc enzo}}

\newcommand{\Ms}{\ifmmode{M_\odot}\else{$M_\odot$}\fi}
\newcommand{\vrms}{\ifmmode{v_{\rm rms}}\else{$v_{\rm rms}$}\fi}

\newcommand{\hh}{H$_2$}

\newcommand{\tvir}{\ifmmode{T_{\rm{vir}}}\else{$T_{\rm{vir}}$}\fi}
\newcommand{\mvir}{\ifmmode{M_{\rm{vir}}}\else{$M_{\rm{vir}}$}\fi}
\newcommand{\rvir}{\ifmmode{r_{\rm{vir}}}\else{$r_{\rm{vir}}$}\fi}

\newcommand{\jj}{\ifmmode{J_{21}}\else{$J_{21}$}\fi}
\newcommand{\flw}{\ifmmode{F_{LW}}\else{$F_{LW}$}\fi}
\newcommand{\kph}{\ifmmode{k_{\rm ph}}\else{$k_{\rm ph}$}\fi}

\newcommand{\zsun}{\ifmmode{\rm\,Z_\odot}\else{$\rm\,Z_\odot$}\fi}

\newcommand{\hi}{H {\sc i}}
\newcommand{\hii}{H {\sc ii}}
\newcommand{\hei}{He {\sc i}}
\newcommand{\heii}{He {\sc ii}}
\newcommand{\heiii}{He {\sc iii}}
\newcommand{\nhi}{\ifmmode{N_{\rm HI}}\else{$N_{\rm HI}$}\fi}

\newcommand{\oiii}{[O {\sc iii}]}

\newcommand{\cii}{C {\sc ii}]}

\newcommand{\nii}{N {\sc ii}}
\newcommand{\civ}{C {\sc iv}}

\def\eps@scaling{1.0}% 
\newcommand\epsscale[1]{\gdef\eps@scaling{#1}}% 
\newcommand\plotone[1]{% 
 \centering 
 \leavevmode 
 \includegraphics[width={\eps@scaling\columnwidth}]{#1}% 
}% 
\newcommand\plottwo[2]{% 
 \centering 
 \includegraphics[width={\eps@scaling\columnwidth}]{#1}% 
 \hfil 
 \includegraphics[width={\eps@scaling\columnwidth}]{#2}% 
}% 

\title[He II Emission in High Redshift Galaxies]{Blue Galaxies: Modeling Nebular \heii{}\ Emission in High Redshift Galaxies}

\author[K. S. S. Barrow]{Kirk S. S. Barrow$^{1}$\thanks{e-mail:
    kssbarrow@stanford.edu}\\
  $^{1}$ Kavli Institute of Particle Astrophysics and Cosmology, Stanford University, 452 Lomita Mall
Stanford, CA  94305-4085\\
}
\pagerange{\pageref{firstpage}--\pageref{lastpage}} \pubyear{2019}

\date{Accepted November 22, 2019. Received November 20, 2019; in original form September 8, 2019}

\begin{document}
\label{firstpage}
\pagerange{\pageref{firstpage}--\pageref{lastpage}}
\maketitle

\begin{abstract}

Using cosmological simulations to make useful, scientifically relevant emission line predictions is a relatively new and rapidly evolving field. However, nebular emission lines have been particularly challenging to model because they are extremely sensitive to the local photoionization balance, which can be driven by a spatially dispersed distribution of stars amidst an inhomogeneous absorbing medium of dust and gas. As such, several unmodeled mysteries in observed emission line patterns exist in the literature. For example, there is some question as to why \heii\ $\lambda 4686$/H$\beta$ ratios in observations of lower-metallicity dwarf galaxies tend to be higher than model predictions. Since hydrodynamic cosmological simulations are best suited to this mass and metallicity regime, this question presents a good test case for the development of a robust emission line modeling pipeline. The pipeline described in this work can model a process that produces high \heii\ $\lambda 4686$/H$\beta$ ratios and eliminate some of the modeling discrepancy for ratios below 3\% without including AGNs, X-ray binaries, high mass binaries, or a top-heavy stellar initial mass function.  These ratios are found to be more sensitive to the presence of 15 Myr or longer gaps in the star formation histories than to extraordinary ionization parameters or specific star formation rates. They also closely correspond to the WR phase of massive stars. In addition to the investigation into \heii\ $\lambda 4686$/H$\beta$ ratios, this work charts a general path forward for the next generation of nebular emission line modeling studies.

\end{abstract}


\section{Introduction and Background}

\hii\  regions are regions in galaxies where a source with a hard spectrum, such as a young stellar cluster, photo-ionizes a large volume of hydrogen in the interstellar medium (ISM).  Compared to regions of neutral gas, \hii\ regions are internally transparent to ionizing radiation and are thus host to chemical species in various states of ionization. These species in turn emit most intensely in the visible and UV portions of the electromagnetic spectrum. Within our own galaxy, these regions appear as hazy, distributed sources of colorful light and were described by classical astronomers as nebulae for their ``nebulous" appearance. Two emission lines, \heii\ $\lambda 4686$ and H$\beta\ (\lambda 4861)$, notably give some \hii\ regions a distinct aqua-blue color, which has come to be broadly associated with evidence of recent star formation by astronomers.

Since \heii\ emission requires the recombination or excitation of singly ionized helium, and thus higher nebular temperatures and a harder spectrum than H$\beta$ emission, the ratio of \heii\ $\lambda 4686$ to H$\beta$ has been used to characterize the temperature of the radiative source of the \hii\ region. Strong nebular \heii\ emission powered by young stars have been identified in extra-galactic \hii\ regions as early as 1979 \citep{1991ApJ...373..458G} with an average \heii\ $\lambda 4686$/H$\beta$ ratio of 2\%. However, emission line diagnostics of galactic spectra have revealed even higher \heii\ $\lambda 4686$/H$\beta$ ratios averaged over their entire ISM, which has been difficult to explain with only contributions from young main sequence stellar populations \citep{2018Galax...6..100O}. For example, the \citet{2001MNRAS.323..887C} modeling technique using \citet{2004MNRAS.351.1151B} parameter space predicted \heii\ $\lambda 4686$/H$\beta$ ratios that were as much as an order of magnitude less than the 1-4\% values observed among low metallicity galaxies ($Z < 0.2\ Z_\odot$)\citep{2012MNRAS.421.1043S}.

Several explanations have been proposed in the literature. Wolf-Rayet (WR) stars and WR galaxies (galaxies with stellar WR emission lines from several contributing WR stars) have been shown to produce strong broad \heii\ $\lambda 4686$ emission features \citep{1998ApJ...497..618S} and could serve as part of an explanation. However, \citet{2000ApJ...531..776G} show that WR features were not present in the spectra in all galaxies with high \heii\ $\lambda 4686$/H$\beta$ ratios and suggest another contributing mechanism like shocks. \citet{2013MNRAS.431..493K} show that shocks in stellar-driven nebulae do have the potential to increase the strength of the \heii\ $\lambda 4686$ line, but their models still underestimated the observed line ratios. 

Since an active galactic nuclei (AGN) produces energetic radiation and \heii\ $\lambda 4686$ emission lines, they have also been explored as an explanation for the discrepancy. AGN-hosting systems indeed have extremely strong \heii\ lines that are as many as ten times the values seen in individual \hii\ regions, but the presence of an AGN does not explain the complete sample of galaxies with high \heii\ $\lambda 4686$/H$\beta$ ratios. \citet{2012MNRAS.421.1043S} show that AGN-powered systems are clearly delineated from purely star-forming high \heii\ emission systems upon examining their place on a \heii\ $\lambda 4686$/H$\beta$ versus [\nii]\  $\lambda 6584$/H$\alpha$ plot. Furthermore, shock-powered, AGN-powered, and star-powered high \heii\ emission systems are all delineated from each other on the vertical axis of a  [\civ]\ $\lambda \lambda 1548,1550$/\cii\ $\lambda 1908$ versus \oiii\ $\lambda 1666$/\heii\ $\lambda 1640$ plot \citep{2017MNRAS.472.2608S}.

Taken together, there is some evidence of an unknown process specifically within star-forming galaxies that produce a larger than expected \heii\ $\lambda 4686$/H$\beta$ ratio. The effect has been shown to be most prominent in systems with lower metallicity as defined as 12 + Log$_{10}$[O/H] < 8.0 in both WR \cite{2010A&A...517A..85L} and non-WR rich systems \citep{2017MNRAS.472.2608S}. One explanation may be that there are more higher mass stars in the initial stellar mass function (IMF) of star clusters powering higher ionization rates. This is roughly in line with the expectation that lower-metallicity stars tend towards higher masses than higher-metallicity stars, but this is in some tension with common stellar population synthesis models that have already accounted for empirical and analytic metallicity relationships in their IMF prescriptions \citep[e.x.][]{1999ApJS..123....3L,2003MNRAS.344.1000B,2010ApJ...712..833C}. Advancements in stellar modeling may address some of tension by uncovering more processes that result in higher levels of ionizing radiation \citep{2016MNRAS.457.4296W}. Recently, sophisticated stellar models have demonstrated that higher mass, lower luminosity binary stars may generate an ionizing continuum comparable to WR stars \citep{2017A&A...608A..11G} for example.

In this work, the latest techniques in radiative transfer are used to determine whether high \heii\ $\lambda 4686$/H$\beta$ ratios can be reproduced with fiducial models of star formation in  sub-solar metallicity galaxies within a cosmological context. Specifically, this work seeks to advance the state-of-the art in the modeling of nebular emission line production in time-dependent, non-equilibrium, non-LTE, inhomogeneous galaxy-scale environments. This serves two purposes. The first is to confirm whether the modeling discrepancy that motivated the search for alternate sources of \heii\ $\lambda 4686$ for the past couple decades persists, and if so, the magnitude of the discrepancy. The second is to provide robust nebular emission line predictions for dwarf galaxies at redshifts ($z$) greater than 4.5, where detailed radiation-hydrodynamic cosmological simulations are most proficient. 

This work contributes to emission line modeling theory, so some effort is spent on first building up the details of the methodology framework as well as in justifying the modeling assumptions in Section \ref{sec:meth}. Then, in Section \ref{sec:resul}, $\lambda 4686$ and H$\beta$ luminosities are modeled from a series of cosmological simulations to try to understand the source and the physical mechanisms that precede their production. Section \ref{sec:diss} focuses on comparisons between model predictions and observations in the literature and makes predictions for observed occurrences of >2\%  \heii\ $\lambda 4686$/H$\beta$ ratios before this work concludes with a summary of findings in Section \ref{sec:con}.

\section{Methods}
\label{sec:meth}

Four suites of simulations are used to stage a post-processed radiative transfer calculation of emission line production. Data from the Renaissance Simulations \citep{2015ApJ...807L..12O,2016ApJ...833...84X} at redshifts 8 and 11.6 are used to test the modeling pipeline and produce a snapshot of observations. That analysis is followed by a time domain analysis on cosmological simulations focused on $\rm{10^{10}\ M_\odot}$ galaxies at $4 < z < 5.5$ to parse the dynamics behind emission line observations. 

\subsection{Cosmological Zoom-in Simulations}

Preliminary analyses focus on snapshots of (5 cMpc/$h)^3$ ``zoom-in" regions of a (40 cMpc/$h)^3$ Renaissance Simulations super volume that was used to stage a suite of high-resolution hydrodynamic cosmological simulations. These employ 9-species chemical networks (\hi{}, \hii{}, \hei{}, \heii{}, \heiii{}, $\rm{e^-}$, $\rm{H^-}$, $\rm{H_{2}^+}$, $\rm{H_{2}}$), radiating metal-enriched and metal-poor star formation routines, as well as supernovae feedback run with the adaptive mesh refinement (AMR) code \enzo{} \citep{2014ApJS..211...19B}. Simulations were run assuming the cosmological parameters: $\Omega_M = 0.266$, $\Omega_{\Lambda} = 0.734 $, $\Omega_b = 0.0449$, $h = 0.71$, $\sigma_8 = 0.8344$, and $n = 0.9624$, which are taken from the 7-year results of the $Wilkinson\ Microwave\ Anisotropy\ Probe$ \citep[WMAP;][]{2011ApJS..192...16L}. 

In zoom-in simulations such as the ones used in this study, the root grid, defined here as AMR level zero, is initialized with pre-refined nested sub-grids around a volume of interest and then allowed to further refine to higher AMR levels within those volumes after reaching a refinement criterion. The Renaissance simulations are run with a (512)$^3$ root grid dark matter resolution and further refined by four AMR levels in static Lagrangian volumes about overdense, underdense, and normal density sections of the full volume. Since each refinement level increases the dark matter particle density by a factor of eight, zoom-in regions attain an effective resolution of (4096)$^3$, which corresponds to a maximum dark matter mass resolution of $2.9 \times 10^4\ \rm{M_\odot}$. Gas particles are allowed to further refine to a higher level upon the condition that gas overdensity reach at least 40 times the mean density of its current subgrid level up to a maximum of the 12th AMR level. Though few cells trigger the 12th level refinement criterion, this corresponds to a spatial resolution of at most 19 pc $(1+z)^{-1}h^{-1}$. 

The underdense ``Void" zoom-in region was chosen for preliminary comparisons with observations because it is the only set from the Renaissance project that probes the end of reionization, due to the relatively lower computational cost of simulating a region with fewer galaxies than the other zoom-in regions. In total there are fifty star-forming galaxies in the Void simulation of which twenty have total halo masses over $10^7\ \rm{M_\odot}$ and stellar masses greater than $10^5\ \rm{M_\odot}$. The most massive galaxy in the Void region has a total halo mass of  $6.81 \times 10^9 \ \rm{M_\odot}$ and a stellar mass of $1.87 \times 10^8 \ \rm{M_\odot}$ at $z=8$.

The average-density ``Normal" region of the Renaissance simulations is also analyzed to provide context between the two overdensities and to provide more examples of larger galaxies. Two simulations of the Normal region were run. The first simulation uses a fiducial \cite{2012ApJ...746..125H} UV background and the second employs a self-consistent UV background derived from the radiation field calculated in the first simulation. This allows the second simulation to more robustly model the important effect of radiative feedback on star and galaxy formation. Therefore, the latest redshift of the second Normal ``BG1" simulation ($z=11.6$) is used as the basis of this work's analysis of the Normal region. At this redshift, there are twenty-five galaxies with total halo masses over $10^7\ \rm{M_\odot}$ and stellar masses greater than $10^5\ \rm{M_\odot}$. The most massive galaxy has a total halo mass of $3.24 \times 10^9 \ \rm{M_\odot}$ and a stellar mass of $6.53 \times 10^7 \ \rm{M_\odot}$. A similar effort to create a self-consistent UV background for the Void region is ongoing and not available to $z=8$ at the time of writing, so a preference for an analysis of a lower redshift takes precedence over using the more self-consistent simulation for that portion of this study.

The third set of zoom-in simulations focus on the evolution of individual halos over time and are initialized with parameters produced as part of the {\sc Agora} collaboration \citep{2014ApJS..210...14K}. After running a dark matter only simulation, {\sc Agora} constructed zoom-in regions that contained the progenitors of halos and were isolated from other halos at $z=0$. For this work, \enzo{} is again used to run two radiation-hydrodynamic simulations of {\sc Agora} zoom-in regions centered on isolated halos that accumulate a total halo mass of $10^{10}\ \rm{M_\odot}$ by $z=0$; one with a violent merger history (10v) and one with a quiescent merger history (10q). Both simulation volumes are (5 cMpc/$h)^3$ and have $\sim$ (1 cMpc/$h)^3$ zoom-in regions with effective root grid dimensions of (2048)$^3$. They are run with cosmological parameters $\Omega_M = 0.3065$, $\Omega_{\Lambda} = 0.6935 $, $\Omega_b = 0.0483$, $h = 0.679$, $\sigma_8 = 0.8344$, and $n = 0.9681$, which are taken from the most recent release of the \citet{2018arXiv180706209P}. They are also run with physics routines that are nearly identical to the Void and Normal simulations, but with finer dark matter resolution ($\sim 1000\  \rm{M_\odot}$) and up to the 14th AMR level relative to the root grid. 10q and 10v achieve up to the 12th AMR level during the redshifts analyzed during this study, which corresponds to a spatial resolution of at best 9.54 pc $(1+z)^{-1}h^{-1}$ or about (2.56 pc)$^3$ per cell at $z=4.5$. A \cite{2012ApJ...746..125H} UV background is applied to both simulations to drive reionization and simulate the presence of nearby galaxies. Outputs at redshifts $4 < z < 5.5$ from two versions of the 10q simulation are used as the basis for most of the analysis in this work.

\subsubsection{Simulation Star Cluster Modeling}

Stars in the cosmological simulations are modeled as radiating, gravitationally interacting particles \citep{2012MNRAS.427..311W} that accrete their mass from the surrounding gas after a formation criterion is triggered. To briefly summarize, star formation from low metallicity gas ($Z < 10^{-5.6}\ Z_\odot$ in Renaissance, $Z < 10^{-7.2}\ Z_\odot$ in 10q and 10v) is managed by a dedicated Population III star particle routine and occurs under the condition that the gas has a converging flow ($\nabla \cdot v < 0 $), that the cooling time is less than the dynamical time, and when both molecular hydrogen (\hh) fractions and gas overdensities are high enough (> $5\times 10^{5}$) to implicate a collapsing metal-free molecular cloud. The final accreted mass of Population III star particles is selected from a Salpeter-like initial mass function (IMF) with a low-mass exponential cut off and is limited to masses between 1 and 300 $\rm{M_\odot}$.  Population III star particles then emit ionizing radiation for a mass-dependent lifetime \citep{2003ApJ...591..288H}, after which they inject metals and mass-dependent supernovae energies \citep{2002A&A...382...28S} into the gas.

Though proper handling of Population III stars is important to model early feedback and metal enrichment of mini-halos at high redshift, the halos studied in this work are effectively entirely dominated by stars formed from metal-enriched gas. These metal-enriched ``Population II" stars are modeled in the simulation as particles representing multi-star clusters with an initial minimum mass of 1000 $\rm{M_\odot}$. Cluster particles radiate ionizing photons for 20 Myr, corresponding to the maximum lifetime of OB stars. In the Void and Normal simulations, Population II clusters emit hydrogen-ionizing radiation, whereas in 10q, and 10v, clusters emit both hydrogen- and helium-ionizing radiation. Population II cluster particles inject energy as supernovae at $\sim$4 Myr and thereafter continually return metals to the gas for the next $\sim$16 Myr as detailed in \citet{2012MNRAS.427..311W}. Importantly, due to the resolution of the simulation, individual starbursts in the 10q and 10v simulations consist of multiple star particles with unique formation times, positions, and velocities, which thus dynamically disperse over time. This is physically distinct from either an instantaneous burst model or models that assume that star clusters at point sources and is a significant modeling strength when trying to untangle the radiative evolution of an \hii\ region.

\subsection{Emission Line Modeling}

The tools used for the modeling of emission lines in this work were primarily developed to overcome the limitations inherent in modeling individual \hii\ regions in an irregularly shaped and inhomogeneous galaxy with multiple episodes of star formation. Since radiative feedback and metal-injection is handled self-consistently by the cosmological simulation, each computational cell contains a distinct and physically motivated gas density, composition, temperature, and momentum. However, since the simulation only includes a rough estimate of single bands of radiation (Lyman-Werner, hydrogen-ionizing, helium ionizing), a more detailed treatment of flux is modeled in post processing to drive photo-ionization calculations.

\subsubsection{Radiative Sources}

One technique for modeling galactic spectra is to sum the spectra of the individual stars or star clusters and perform a photoionization calculation using ionization parameters and an estimate for the density profile and composition of the galaxy and producing and emission line a look-up table with the {\sc Cloudy} photoionization solver \citep{2013RMxAA..49..137F,2017RMxAA..53..385F}. This technique is used for Flexible Stellar Population Synthesis \citep[FSPS;][]{2010ApJ...712..833C} handling of emission lines \citep{2017ApJ...840...44B}, which was found to be consistent with predictions of various line ratios. However, since the source of the observed \heii{}/H$\beta$ lie ratios are matter of some mystery and debate, it may become important to carefully consider the effect of the entire stellar spectrum and the various combinations of density, flux, and composition that might exist inside a single galaxy.

To explore the effect of a full stellar spectra on photoionization calculations, emission from Population II and Population I stars are modeled using continuum FSPS solutions, which use 21 metallicity isochromes and stellar age to determine the mass function averaged spectra between $\sim$94\AA\ to $\sim$10$^7$\AA\ at 8,000 wavelengths. Though FSPS is parameterized to produce a continuous age-luminosity relationship, calculations are accelerated by using pre-computed solutions for two hundred equally spaced cluster ages from 0-20 Myr and a further two hundred ages from 200 Myr to 2 Gyr to match simulation star particles to stellar spectra of the nearest age and metallicity isochrome. Thus, almost all the bolometric luminosity of stellar clusters can be estimated with enough resolution to capture more subtle fluctuations in the ionizing and UV spectrum, especially during the shorter lifetime of massive stars within clusters.

To explore the emission line-resolution relationship, star clusters in the 20 most massive galaxies in the Void simulation are matched to FSPS spectra and a {\sc Cloudy} photoionization calculation is performed on the encompassing computational cells using their volume, density and metallicity. Here, calculations run into several computational and practical limitations. The virial radius of the largest galaxy in the Void contains over 40,000 star particles and about one million cells. In the most massive halos in the Void and Normal simulations, even individual cells at the highest AMR levels may contain dozens or even more than a thousand star particles. To prioritize computational effort on advancing polychromatic modeling techniques, star particles in halos like these are grouped as to limit the number of sources to less than 5,000. In halos with more than 5,000 sources, starting with the cell with the most interior star particles, the brightest fifty sources are left ungrouped and the remaining particles are sorted into up to twenty spatially consistent subgroups. For each of these groups, the luminosity weighted center and the summed spectra is used for subsequent photionization post processing calculations. The process continues with the next most populated cell until the number of individual and grouped star particles is less than or equal to 5,000. In practice, this results in only a tiny discrepancy in the flux distribution of the halo and protects calculations against running into memory and wall-clock limitations. This process is not employed with the 10q and 10v simulations as they contain halos with no more than 3010 star particles during the time intervals analyzed in this study.

In lieu of performing as many {\sc Cloudy} calculations as there are cells per galaxy, one method for determining the emission line production within all the cells would be to produce a look-up table with values for all the many different combinations of intensity, spectra, composition, and density. However to be useful, the resulting table would be similarly require more than a million {\sc Cloudy} calculations. \citet{2019arXiv190101272K} address this by applying a machine learning algorithm based on a training sub-set of such a look-up table and are able to predict emission line strengths in the rest of the table with decent accuracy at higher wavelengths (>50 $\mu$m), however with larger errors (5-20+\%) at the optical and UV wavelengths that are the focus of this study.

To more precisely calculate optical and UV emission line strengths, an efficient method for accurately modeling the photon intensity within halos is formulated for this study by conducting two separate investigations. First, a generalized model for the multi-wavelength radiative intensity distribution throughout interstellar medium (ISM) of an arbitrary galaxy is formulated. Then a separate analysis of radiative flux near to and within clusters is conducted to determine a model for regions where gas is brightly illuminated.

\subsubsection{Radiative Intensity Throughout the Interstellar Medium}

\begin{figure*}
\begin{center}
\includegraphics[width=\textwidth]{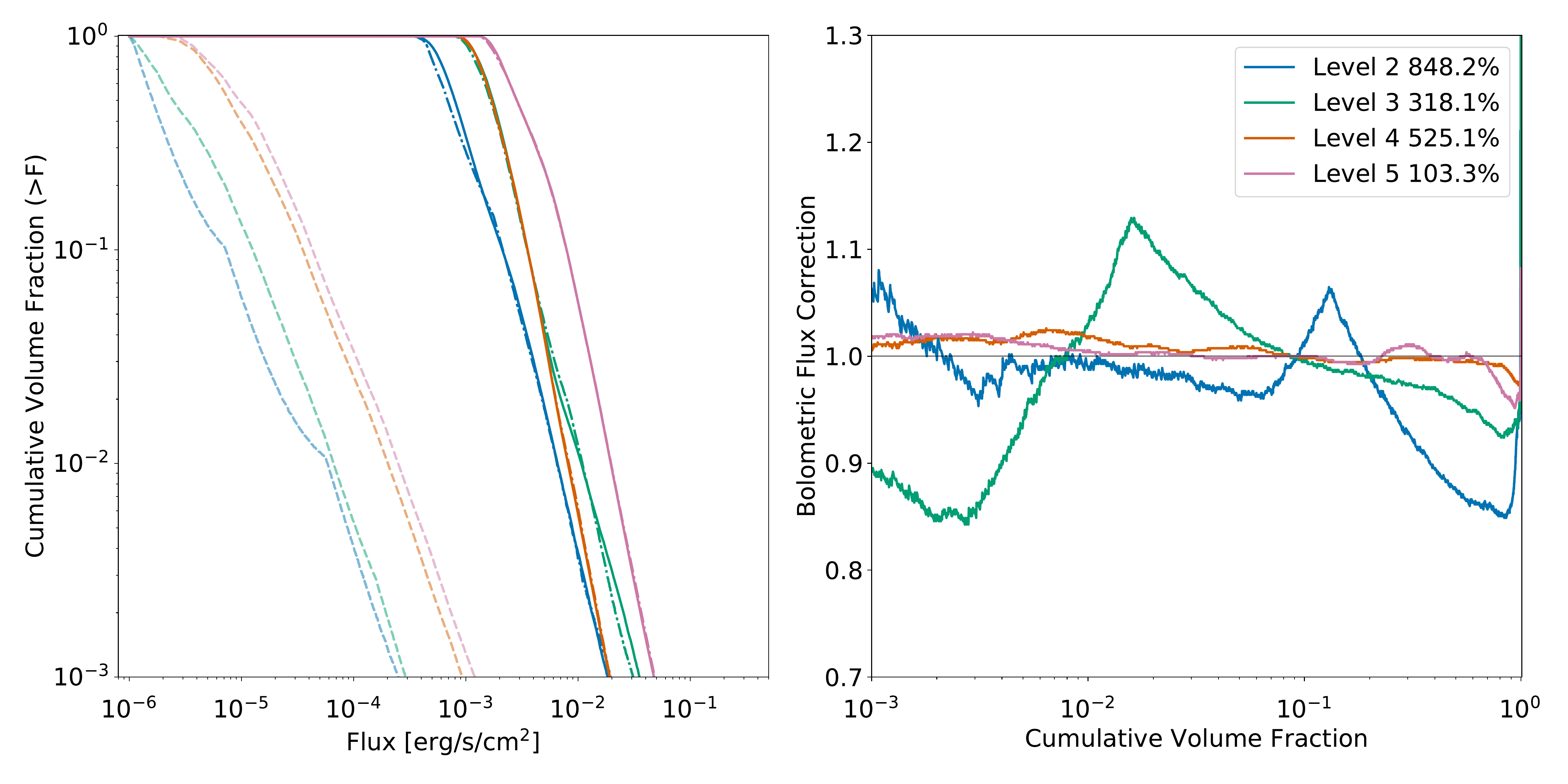}
\caption{Plotted are a comparison between different flux modeling techniques showing that the method in this work is only prone to small <5\% errors as compared to other assumptions and models that might produce order of magnitude errors. Left: Cumulative volume fraction versus flux at 1148 Myr after $z=100$ for the main halo in the 10q simulation using the {\sc Caius} pipeline. Dashed lines demonstrate the flux distributions when only considering stars within cells sized to the second (blue), third (green), fourth (orange), and fifth (purple) AMR level of the simulation. Solid lines indicate the flux distribution when considering all sources within the halo. Dot-dashed lines show the flux estimated with the method used in this work. Right: The difference between estimated and calculated fluxes are plotted as a function of cumulative volume fraction for each of the AMR levels. For both plots, AMR levels two and three only include cells that have interior star particles and AMR levels four and five include both cells cells with interior star particles as well as adjacent cells. The percentages displayed in the legend denote the percent of the virial volume covered by the calculation. Modeled flux estimates at higher levels generally produce smaller errors with respect to calculated galactic flux distribution. }
\label{fig:fluxcorrection}
\end{center}
\end{figure*}

One key limitation of photoionization calculations using {\sc Cloudy} is that users need to choose between a constant intensity mode or modes that assume spherical symmetry. A given cell in a simulation may however be illuminated by an unregularized distribution of radiative intensity. Properly accounting for this distribution is important for photo-ionization calculations that depend on photon density like the production of \heii{} emission lines.

Using mean galactic FSPS spectra, a temperature- and wavelength-dependent mass attenuation coefficient look-up table is produced to determine the spectra in each cell. Using {\sc Cloudy}'s constant temperature model and the mean galactic gas density, two hundred logarithmically spaced temperatures between 316 K and 10$^7$ K are used with the mean FSPS-derived galactic continuum spectra from the star particles to estimate gas mass attenuation coefficients at 50,000 wavelengths from the infrared to the soft X-ray photon energy regimes. Then, rays are traced through the simulation from each star particle towards the center of each cell to calculate the density- and temperature-dependent polychromatic absorption of starlight impinging each cell. Rays from sources exterior to the cell are terminated on the cell boundary and sources interior to the cell are left unattenuated to allow a subsequent photo-ionization calculation to determine absorption self-consistently.

From the attenuated bolometric luminosity of each external star particle and the unattenuated bolometric luminosity of any internal star particles, the flux distribution within a subject cell is determined by calculating the intensity at (50)$^3$ equally distributed points within the volume of the cell. Then, by assuming a power-law distribution of flux versus volume fraction, an equivalent spherical approximation of the flux is parametrized based on the ratio of the highest, $\rm{F_{99}}$. to lowest, $\rm{F_{01}}$, percentile of flux within the cell, ${\rm F_{99}/F_{01}} = [(1+b)/b]^2$, the mean path through the cell, $k\Delta x =a$, and the volume of the cell, $(\Delta x)^3 =(a/k)^3$, where $a$ and $b$ are defined in this manner to simplify expressions and $\Delta x$ is the width of the cell. Thus, this spherical region has an inner and outer radius given by

\begin{equation}
R_i = ab,\ R_o = a(b+1),
\label{eq:RiRo}
\end{equation}

a covering fraction given by

\begin{equation}
f = \frac{3}{4 \pi k^3(3b^2+3b+1)},
\label{eq:covering}
\end{equation}

and an equivalent bolometric luminosity given by

\begin{equation}
L = 4\pi {\rm F_{99}}(ab)^2.
\label{eq:lum}
\end{equation}

The value of $k$ is difficult to determine in an averaged sense since rays from sources inside and outside the volume take a continuum of paths through a cell and thus experience variety of optical depths. Choosing to focus these calculations on interior sources due to their higher fractions of ionizing radiation, the geometric value of $k\approx 0.640396$ is used for Equations \ref{eq:RiRo}, \ref{eq:covering}, and \ref{eq:lum}. 

Taken together, this treatment therefore accounts for the effect of the density substructure on the radiation field down to the finest AMR level. Figure \ref{fig:fluxcorrection} shows an example of the error resulting in using this spherical treatment of radiative transfer on cubic regions, shown as dot-dashed lines in the left-hand plot, as compared to actual flux throughout the same volume, shown as solid lines. The right-hand plot of Figure \ref{fig:fluxcorrection} shows that this treatment models bolometric flux to within 4\% of the true value within 99.9\% of the volume if a sufficiently high AMR level is used. Also shown in the left-hand plot of Figure \ref{fig:fluxcorrection} is the flux distribution when assuming that each star-encompassing cell is independently illuminated by interior sources (dashed lines), which inaccurately estimates the true flux distribution by up to two orders of magnitude and is henceforth abandoned as a technique. Indeed, a separate test showed that only considering ionizing radiations originating from within individual cells underestimated H$\beta$ emission by a factor of two and results in higher than physical line ratios. In addition, a model where flux versus volume is calculated under the assumption that all galactic emission comes from a single source embedded in gas with the average density of the galactic halo tended to overestimate the flux at every value of cumulative volume fraction by an order of magnitude in some cases. This shows that the photon densities within halos is not robustly approximated by averaging the stellar and gas compositions of galaxies. 

Since only cells encompassing stars or adjacent to cells encompassing stars are included in the calculation, the fraction of the virial volume covered by explicit photo-ionization calculations decreases with the maximum AMR level used as shown in Figure \ref{fig:fluxcorrection}'s legend. To balance computational demands with a desire for completeness, the third or fourth AMR level of cell is used for analysis of the Renaissance simulations, depending on the number of sources and the physical extent of a halo. In the Void simulation, this means that emission lines are explicitly calculated in the interstellar medium out to no less than 700 pc and up to 1700 pc from every source compared to a maximum viral radius of 8,070 pc for the largest halo. For the individual galaxy simulations (10v and 10q), the fifth AMR level is used, and calculations typically cover the virial volume. Therefore, effectively all the starlight-induced nebular emissions from inside the star-forming regions out through the circum-galactic medium (CGM) is accounted for by this treatment, and thus the emission lines are calculated as the result of the cumulative effect of light from each of up to 5,000 sources in each halo as well as the attenuating effect of a multi-phase ISM throughout the virial volume. It should also be noted that the AMR levels described here are simply the voxel sizes used for the {\sc Cloudy} photo-ionization calculations and the determination of the radiation properties in each voxel employ ray-tracing that utilizes the best resolution available in the simulation.

\subsubsection{Radiative Intensity Within Stellar Clusters}

To determine radiative intensity in the intra- and circum-cluster medium, a toy model of a star cluster is created. In their review, \citet{2010ARA&A..48..431P} note that star clusters have density profiles that can be described by the EFF profile

\begin{equation}
\rho(r) = \rho_0\left(1+ \frac{r^2}{r^2_c}\right)^{-(\gamma +1)},
\end{equation}

where $\rho(r)$ is the three-dimensional brightness density, $\rho_0$ is the peak central density, $r_c$ is a characteristic radius, and $\gamma$ is an exponent that describes the power law decay in observed surface brightness outward from the center of the cluster. For young clusters, $\gamma$ typically takes values  greater than the lowest possible value of 2 and less than the \citet{1911MNRAS..71..460P} estimate of $\sim$4, so a value of $\gamma = 3$ is adopted for the toy model. Observed young massive clusters (YMCs) in the Local Group listed by \citet{2010ARA&A..48..431P} show that values of $r_c$ that grow as the clusters age from a few tenths of a parsec for the youngest clusters to up to 2 parsecs for clusters that are 10 Myr old. The latter value is adopted for the toy model to accentuate the geometric effect of an extended cluster on the radiative intensity profile. Finally, peak densities of approximately 15 stars/pc$^3$ match observed cluster masses and the outer effective radius is chosen to be 20 pc. Thus, the positions of 27,867 star systems are assigned randomly by drawing from the ensuring radial density distribution and a uniform distribution about the surface a sphere to populate the toy cluster.

Masses for each star are drawn from a \citet{2003PASP..115..763C} IMF for star systems, normalized such that the profiles for sub-solar mass systems and super-solar mass systems form a continuous distribution. The resulting mass of the cluster is between 8000 ${\rm M_\odot}$ and 9500 ${\rm M_\odot}$, depending on the seed used for the random draw of the IMF, which usually includes a handful of OB-type stars. The luminosity of each star system was assigned based on a broken power law, which is equated to stellar mass loss through a power law relationship. This model revealed that flux versus volume distribution derived from treating extended star clusters as single sources was functionally equivalent to treating each star as an individual source. Generally, the difference between the two models is that the interior of the cluster would have a slightly lower peak radiative intensity and a higher radiative intensity in the periphery of the cluster in the many-star model, but the difference is slight. By disregarding the highest percentile of the flux in the definition of $b$, an overestimation of flux from a single source model is avoided.

It is further assumed in the simulation that star particles remain coherent over time, which is at some odds with real-world observations that most stars in galaxies in a distributed stellar field. In this study of emission lines that require high rates of ionizing flux, the sources that affect our measurements are heavily biased towards bright young sources and stellar clusters, so results are less sensitive to this assumption than studies of lines that are produced in lower-luminosity regimes. Furthermore, the quantity of star particles in the simulation lends to a more even distribution of sources than lower-resolution treatments of star formation. Thus, for this model, the approximation of stellar clusters as point sources is likely appropriate. Nonetheless, studying the effect of distributed versus coherent stellar populations deserves some examination. Special attention is paid to analysis of multi-star particle massive cluster formation in this study, which provides a good proxy for cluster dispersion dynamics and the effect of emission from \hii\ regions without modeling each individual star. 

\subsubsection{Dust Corrections and Emission Line Scattering}

Dust and gas attenuation are considered for determining the photon densities throughout halos with this method, which results in more robust predictions for emission line strengths. However, in real galaxies, emission lines are further affected by attenuation and scattering in gas and dust as photons travel towards observers.  Dust corrections to intrinsic emission line strengths are estimated using a ray-tracing Monte Carlo technique using a modified version of {\sc Hyperion} \citep{2011A&A...536A..79R} similar to methods used in \citet{2018NatAs.tmp..126B} and \citet{2018MNRAS.474.2617B}. Temperature- and metalliticty-dependent gas extinction, albedos, and emissivities are combined with a metallicity-scaled \citet{2003ARA&A..41..241D} model for dust. The differences in continuum emission are used to similarly attenuate intrinsic emission line strengths. To test the importance of this effect, dust-corrected outputs from the Normal and Void region were compared to uncorrected outputs. Gas and dust corrections showed a minor effect on the \heii\ $\lambda 4686$/H$\beta$ line ratios (<8\%) and therefore for outputs of the 10q and 10v simulations, only intrinsic line strengths and ratios are reported to steward computational resources.

In addition to continuum gas and dust extinction, some emission lines have high optical depths (e.g. Mg {\sc ii} $\lambda 2803$) due to the presence of photo-sensitive species in the gas so a galaxy-wide photoionization balance may need to be conducted to properly account for absorption, scattering, and redistribution of emission lines as photons travel from \hii\ regions towards observers. To test for this, optical depths for every emission line are recorded during the {\sc Cloudy} calculations of intrinsic emission line strength in each cell.  For the case of the emission lines studied in this work, all line optical depths are less than 0.001 within 5\AA\ of line center along any ray within the virial radius of every halo analyzed. Therefore, secondary photon interactions outside of the initial determination of line luminosities are effectively non-existent and intrinsic line strengths after a cell-scale photoionization balance are reported without further accounting for line attenuation.

In summary, a polychromatic attenuation of incident stellar light from multiple sources and a model for inhomogeneous radiative intensity within cell were found to be critically important to the photoionization balance and the modeling of emission lines. On the other hand, grouping some of the stars in a cluster into point particles and ignoring dust attenuation as well as scattering were not inappropriate or inaccurate simplifications for the calculation of the emission lines for this study.

\section{Results}

\label{sec:resul}

\begin{figure}
\begin{center}
\includegraphics[width=\linewidth]{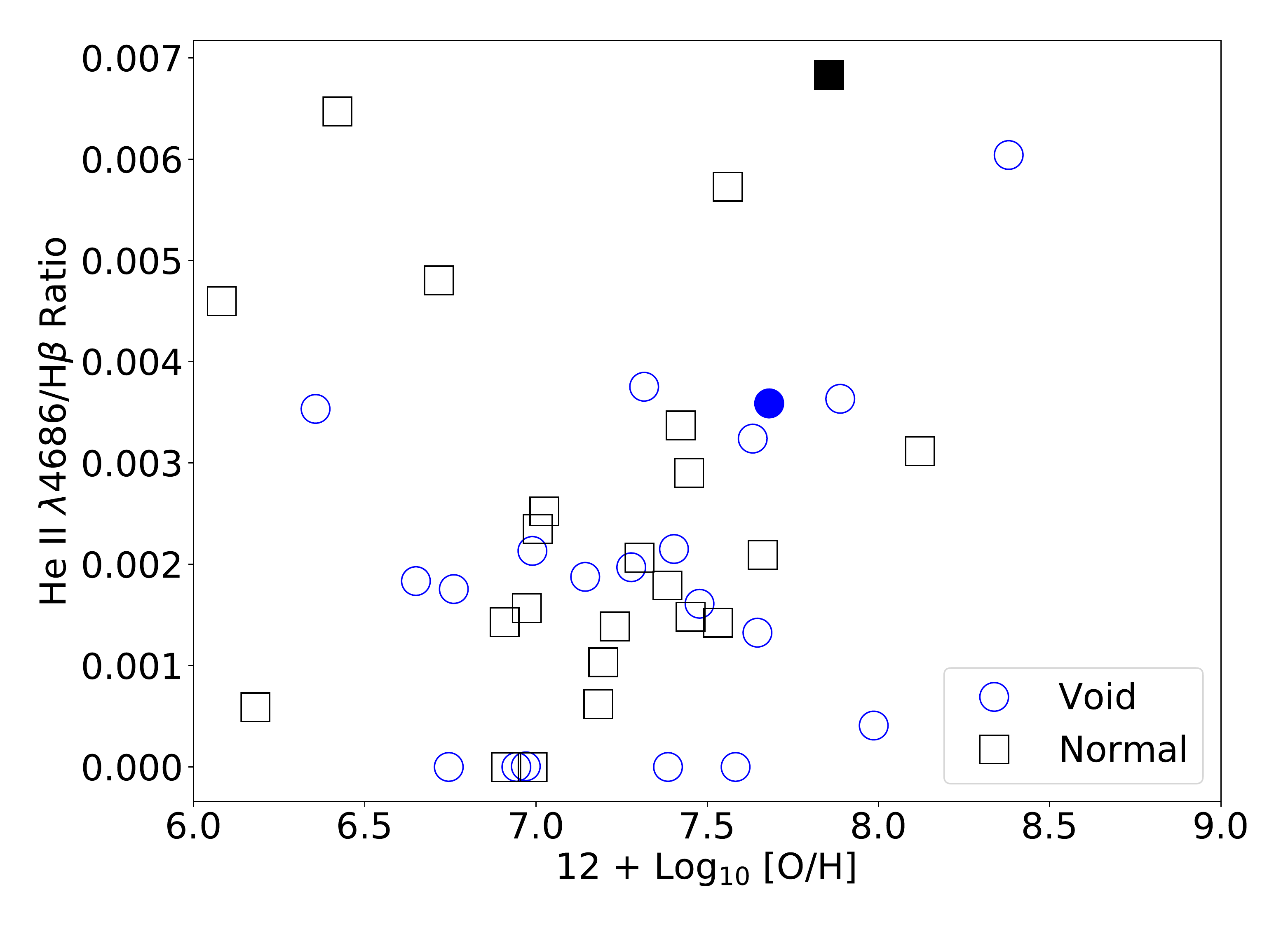}
\caption{Dust-corrected nebular \heii \ $\lambda 4686$/H$\beta$ emission line ratios for galaxies in the $z=8$ Void and $z=11.6$ Normal Renaissance simulations. Solid markers indicate strong \civ\  emission (EW > 0.3). There is not a discernible relationship between metallicity and these line ratios, though the high two values are clustered at 12 + Log$_{10}$[O/H] = 8.0.}
\label{fig:renheii}
\end{center}
\end{figure}

The Void and Normal Renaissance Simulations offer an initial insight into the high-redshift occurrence of galaxies with high \heii \ $\lambda 4686$/H$\beta$ emission line ratios (HeHR). The combined sample is small, but within the confines of our modeling technique, Figure \ref{fig:renheii} shows that the occurrence rate of HeHR greater than 0.7\% is low for (5 cMpc)$^3$ regions of Reionization-era dwarf galaxies. Incidence of strong \civ\ $\lambda 1548$ (equivalent widths > 0.3) is also shown as filled markers to examine whether \civ\ emission is correlated to HeHR as was examined observationally by \citet{2017MNRAS.472.2608S}. Only one galaxy each in the Void and Normal sample exhibited strong \civ\  emission. Unfortunately, both Renaissance samples have low averaged ISM metallicities owing to their limited star formation histories and so inferences on HeHR mechanisms in gas with metallicities of 12 + Log$_{10}$[O/H] > 8.0 are not forthcoming. There is some evidence of a slightly lower ratios at lower metallicity that might be completely attributable to small sample size noise, though this trend might hint at a process that limits the ratio at low metallicities. The Void sample also had generally lower HeHRs than the Normal sample, which might imply a redshift dependence, though again both samples were generally entirely in line with observations of the local Universe and the small sample size makes inferences difficult.

While looking at individual snapshots of simulations is somewhat similar to telescope observations and can useful for making inferences about the expected number observations, it does not yield new physical insight unless one happens to capture an anomalous process or signal by chance. However, studying the evolution of a single galaxy over several hundred million years is much more likely to capture a new process in action. Since $\sim 20$ Myr is the interaction time between young massive stars and \hii\ regions, it should be noted that analysis of longer simulation time intervals tends to give results that could not be used to interpret the frequency or strength of individual nebular emission line processes. In this study, emission line production in the 10q simulation is first examined at time intervals of 1.84 Myr from 1.03 to 1.32 Gyr after the Big Bang. 

\begin{figure*}
\begin{center}
\includegraphics[scale=.30]{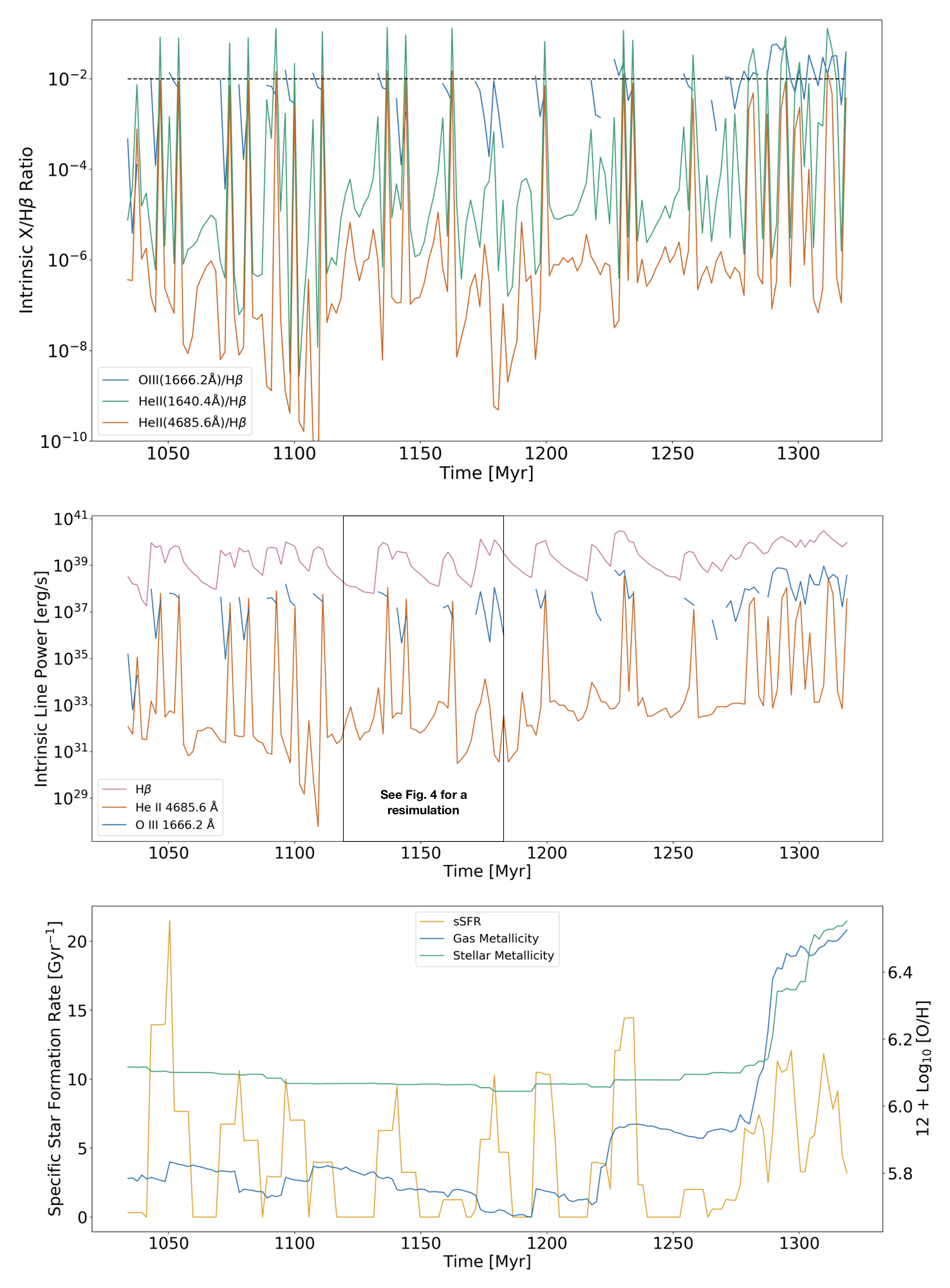}
\caption{Top: Time series of emission line to H$\beta$ ratios for the main halo of simulation 10q plotted over time. Immediately preceding each peak in \heii \ $\lambda 4686$/H$\beta$ is a trough, implying that H$\beta$ peaks before \heii \ $\lambda 4686$. \oiii  \ emission lines can be seen tracing the duration of star formation. The dashed black line indicates ratios of 1\%. Middle: Time series of emission line luminosities over time. Bottom: Specific star formation rate (yellow, left vertical axis) versus time and both gas (green) and stellar (blue) metallicity (right vertical axis) versus time.}
\label{fig:hbratios}
\end{center}
\end{figure*}

%\begin{figure*}
%\begin{center}
%\includegraphics[width=\textwidth]{plotlines3_0.png}
%\caption{Top: Time series of emission line luminosities over time. Bottom: Specific star formation rate (yellow, left vertical axis) versus time and both gas (green) and stellar (blue) metallicity (right vertical axis) versus time.}
%\label{fig:hblums}
%\end{center}
%\end{figure*}

\begin{figure}
\begin{center}
\includegraphics[width=\linewidth]{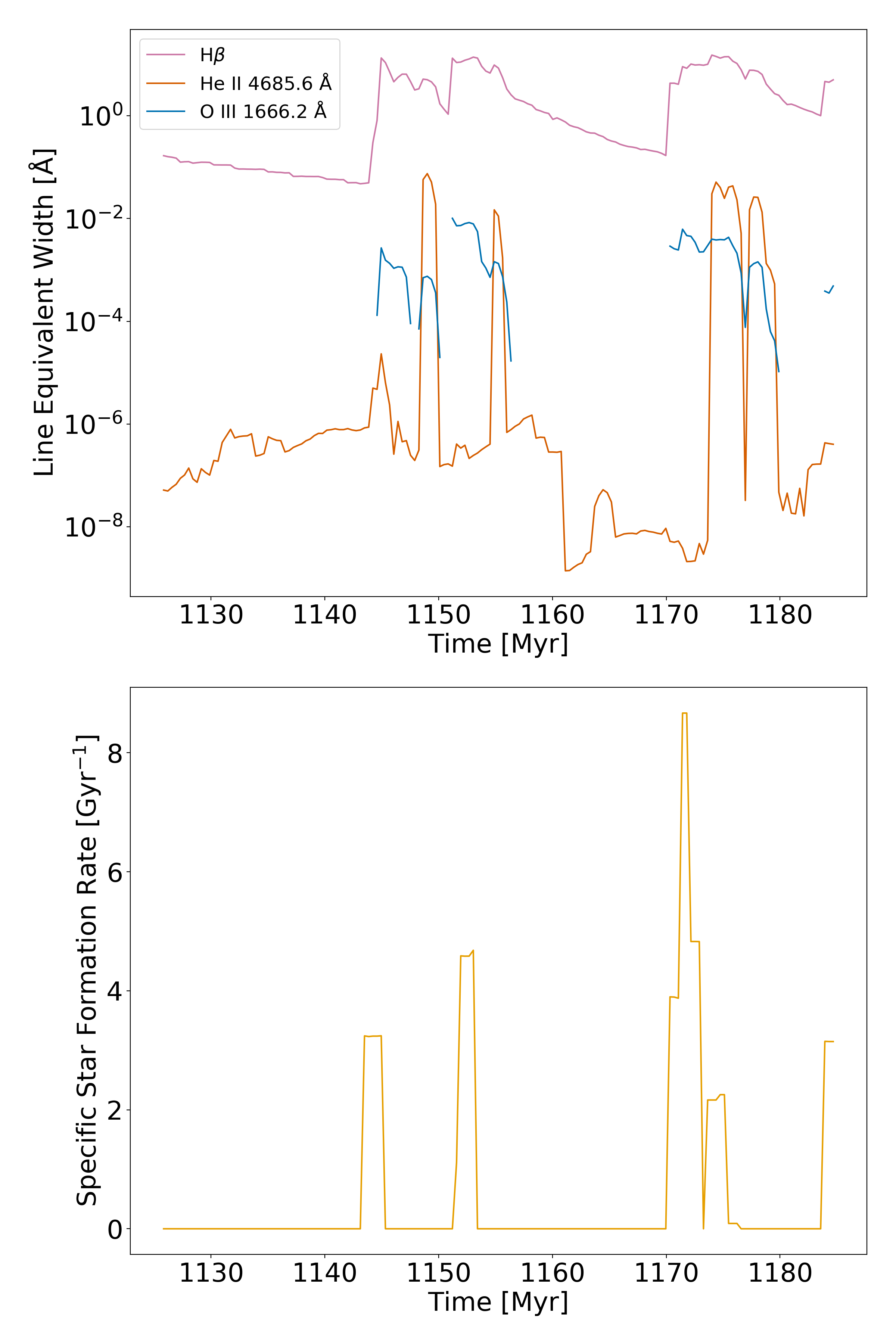}
\caption{A resimulation of the data plotted in Figure \ref{fig:hbratios}, focusing on a shorter time period with a smaller timestep interval (368 kyr). The top plot shows line equivalent width in \AA\ versus time and the bottom plot shows specific star formation rate versus time. The first peak in \heii \ $\lambda 4686$ emission occurs amidst declining H$\beta$ emissivity and thus produces a strong HeHR peak.}
\label{fig:hblums2}
\end{center}
\end{figure}

\begin{figure*}
\begin{center}
\includegraphics[width=\textwidth]{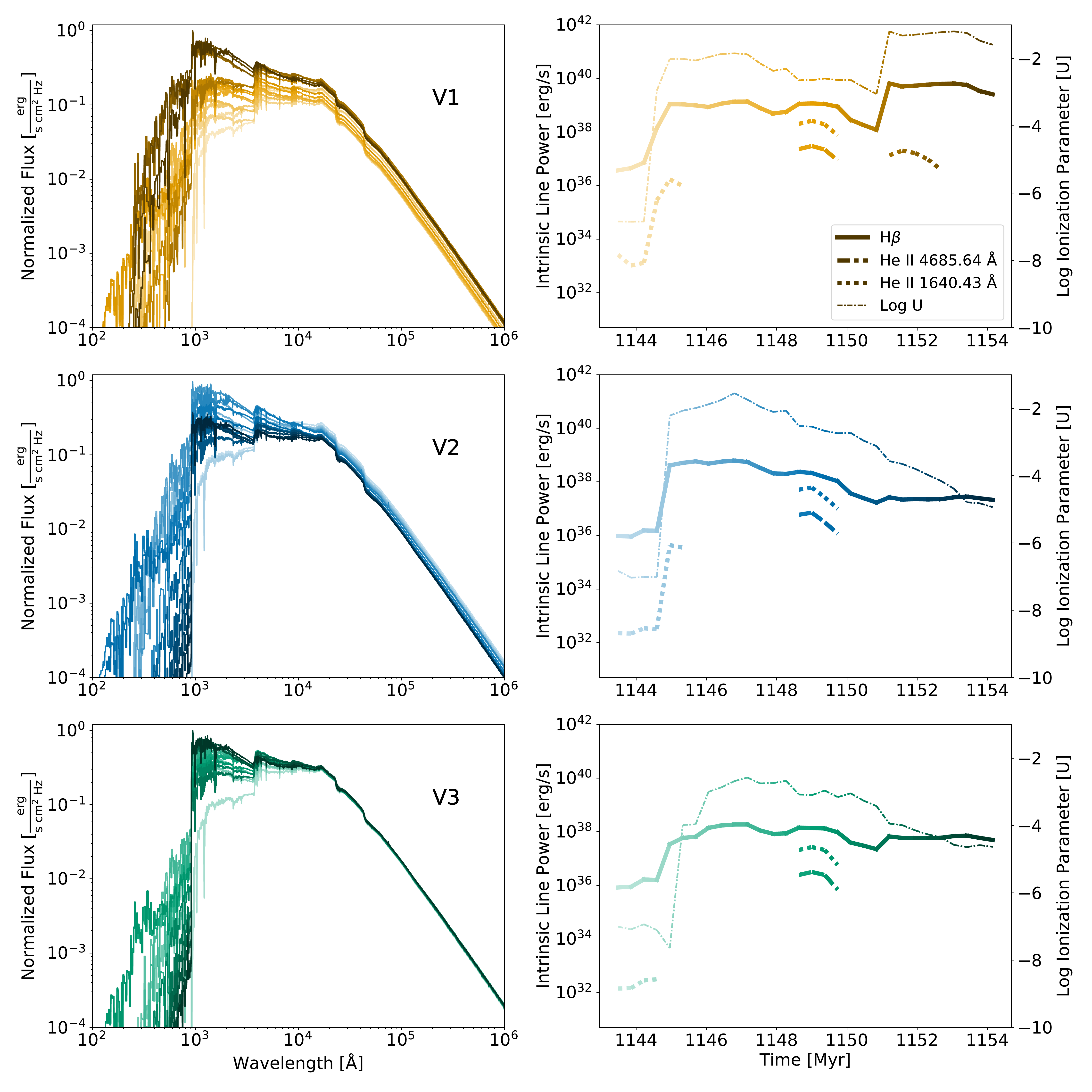}
\caption{The relationship between the ionization parameter, spectrum hardness, and emission line production are shown for the first 12 million years after a starburst. Left Column: Flux versus wavelength in study volumes V1 (orange), V2 (blue), and V3 (green) normalized to the maximum value plotted as time evolves, with darker shades representing later times.  Right Column:  H$\beta$ (thick solid line), \heii \ $\lambda 4686$ (thick dash-dotted line), \heii \ $\lambda 1640$ (thick dashed line) emission (left vertical axis) versus time in V1, V2, and V3 with the same time evolution shading pattern used in the left column. Also plotted are the log ionization parameters (thin dashed line, right vertical axis) for the study volumes using the same color scheme. \heii \ $\lambda 4686$ emission peaks for a short period of time when the spectrum hardens.}
\label{fig:threecells}
\end{center}
\end{figure*}

The top plot of Figure \ref{fig:hbratios} shows emission line ratios within a single halo during this time interval. Therein, HeHRs peak in short, episodic bursts lasting less than 2 Myr, evidenced by each peak occupying a single data point at this temporal resolution. Each burst is also associated with examples of \oiii\ $\lambda 1666$ ratios of a few percent, peaking at their onset a few million years before HeHR peaks.  Additionally, immediately before each HeHR peak, the HeHR falls to minima, which is suggestive of an underlying time-dependent process. The bottom two plots of Figure \ref{fig:hbratios} partially elucidate these phenomena. As one might expect, increases in nebular emission line luminosity are fueled by the presence of young stars and are therefore related to the specific star formation rate (sSFR). Here, sSFR is displayed instead of star formation rate (SFR) because sSFR more intuitively captures the impact of a starburst event on a halo-wide line ratio. Halo-wide H$\beta$ emission luminosity rises almost immediately as stars form and begins to taper off until the next episode of star formation. Since the peak \heii \ $\lambda 4686$ emission luminosity falls within a tight range between 10$^{37}$ and 10$^{38}$ erg/s after each starburst captured in Figure \ref{fig:hbratios}, the galaxy-wide HeHR ratio is mostly modulated by the timing of the two and a half orders of magnitude decay of H$\beta$ emission luminosity on a 10 Myr time scale in the absence of new young stars. When a second starburst occurs within this time frame, H$\beta$ emission luminosity increases and subsequent galaxy-wide HeHRs are therefore muted.

The relative evolution of \oiii\ emission is an important diagnostic that may lead to inaccurate conclusions about the nature of \hii\ regions, shocks and host galaxies. \oiii\ $\lambda 1666$ emission is used to delineate between different sources of high HeHR ratios so it is instructive that luminosity of that line generally also traces star formation, but not as strictly as shown in the middle plot of Figure \ref{fig:hbratios}. After the \oiii\ ratio peaks, it gradually decreases and, in most cases, falls off entirely by the end of the short HeHR peak. These relationships mostly break down at the end of period as the interval between peaks shortens. During that phase, the \oiii/H$\beta$ line ratio is sustained at its highest level and the HeHR peaks are less pronounced, but longer-lived. \oiii\ emission will be explored in more detail in forthcoming studies.

After the 1,275 Myr point, the density-weighted mean gas metallicity rises towards 12 + Log$_{10}$ [O/H] = 6.0 and a period of continuous star formation commences as shown in the bottom plot of Figure \ref{fig:hbratios}. The lower HeHR peaks during this period are explained by sustained high H$\beta$ emission contributions from multiple \hii\ regions. Indeed, the HeHR only exceeds 1.5\% when star formation is intermittent and H$\beta$ therefore has time between starbursts to slope downward. Accordingly, even though the total luminosity of the \heii\ $\lambda 4686$ is greater during high sSFR episodes such as the one that occurs at $\sim$1,225 Myr years after $z=100$, the HeHR peak is higher during a relatively moderate burst with a sSFR less than 10 Gyr$^{-1}$ at $\sim$ 1,150 Myr.

Because each HeHR peak was only captured by a single data point in Figure \ref{fig:hbratios}, the maximum and duration of HeHR emission is unclear. To further examine the HeHR peak at $\sim$ 1,150 Myr, the period from 1,125 to 1,185 Myr is re-simulated with 368,000 yr intervals between simulation snapshots. It should be noted that a re-simulation does not exactly recreate the conditions of the initial simulation due to the parallelization strategy used in \enzo\ and consequentially, the timing of star formation episodes is slightly perturbed. Nonetheless, the resimulation captures a HeHR peak greater than 1.7\% at $\sim$ 1,150 Myr as well as two subsequent peaks as shown in Figure \ref{fig:hblums2}. The three bursts analyzed in Figure \ref{fig:hblums2} would suggest that \heii\ $\lambda 4686$  EW peaks are relatively flat for a period of about 1.5 Myr. As such, the expectation is that the frequency of peaks shown in Figure \ref{fig:hbratios} is fairly representative of both the occurrence rate and intensity of HeHR peaks that occur during the simulated period.

There are no Population III (metal-poor) stars in the halo during the intervals analyzed, so the production of \heii\ $\lambda 4686$ is solely powered by regions around moderately metal-enriched stars. Furthermore, peaks the HeHR tend to lag star formation by a few Myr, which is similar to results from \citet{2012MNRAS.421.1043S}. The time lag is likely due to WR stars in the FSPS spectra. However, to more thoroughly explain the timing, intensity, and duration of the 1,150 Myr HeHR peak, the incident spectra used to drive the photoionization calculation is more closely examined in Figure \ref{fig:threecells}. 

The three rows of Figure \ref{fig:threecells} compare three $\sim$($400$ pc)$^3$ volumes of gas that emit the strongest \heii\ $\lambda 4686$/H$\beta$ ratio during the interval. To follow a consistent gas volume over a period of several million years, the mass-weighted velocity of the gas is used to predict the location of the volume in the next simulation snapshot and the closest cell to the projected location is analyzed and plotted. The first volume, V1, plotted in shades of orange, is close enough to two bursts of star formation to receive a higher cumulative flux of ionizing radiation near the end of the interval. This demonstrates how the flux-matching method can result in a much higher photon density in star forming regions and reinforces the need for this type of modeling when determining the strength of emission lines. The other two volumes, V2 plotted in shades blue and V3 plotted in shades of green, are far enough away from the second burst that they are shielded from the additional ionizing flux, which illustrates the impact of multi-wavelength ray-tracing and absorption modeling.

For the plots in the right column of Figure \ref{fig:threecells}, the ionization parameter in a cell is estimated from the spectra and the hydrogen density of the gas using the following prescription. The density of ionizing photons is a function of the mean ionizing frequency, $\left\langle \nu \right\rangle$, calculated directly from the incident spectra. Given that the photionization calculations are on a spherical sector with an inner and outer radius, the mean flux is related to $1/r_{f}^2$, where the square of the radius representing mean flux, $r_{f}^2$, is 

\begin{equation}
r_{f}^2 = \frac{1}{3}(R_o^2+R_oR_i + R_i^2)
\end{equation}

by simple geometric relations in terms of variables introduced earlier in this work. The mean photon flux is therefore approximately 

\begin{equation}
n_{phot} = \frac{Lf}{4\pi r_{f}^2 \hbar \left\langle \nu  \right\rangle dr },
\end{equation}

where $dr$ is one light second. Finally, $U \equiv n_{phot}/n_{H}$ where $n_{H}$ is the average hydrogen density of the cell \cite{2012ApJ...757..108Y}. This estimate of the ionization parameter systematically overestimates the average number of ionizing photons within the cell because it does not take into account absorption. Likewise, because it is an average, the inner rim of the region representing the most illuminated region of the cell will experience an ionization parameter higher than $U$ by construction. Thus, some care is warranted when comparing the ionization parameter in this work to models using constant intensities or strictly spherical geometries around a single ionizing source. 

The starburst at around $\sim$1,143 Myr produces 48 star particles totaling $6.8 \times 10^5\ \rm{M_\odot}$ inside a 21 pc-wide volume. Importantly, each particle has a different creation time and the evolution of the resulting \hii\ region is \textbf{not} equivalent to a model assuming an instantaneous burst. Over the course of 10 Myr, the dispersion of the star particles grows from 2.5 pc to 241 pc or about as much as 0.4\% of the virial volume of the main halo.  Study volumes V1, V2, and V3 all host star particles from the starburst throughout the interval and all three regions experience an initial increase in Log $U$. However, the rise in the ionization factor in V3 is more gradual than in the other two regions due to dynamics of the gas and mass loss. The hydrogen nucleon density of V3 drops from approximately 16 nuclei/cm$^3$ at 1,145.6 Myr to less than 3.4 nuclei/cm$^3$ by 1,149.4 Myr and  V3 is also initially the furthest from the starburst, but drifts closer to the luminosity-weighted center of the cluster by 1,146.2 Myr. 

Emission line production in all three regions exhibits the same pattern in response to the starburst. \heii\ $\lambda 1640$ and H$\beta$ emission grows in concert with the initial jump in the ionization parameter at 1,144.6 Myr. Then, \heii\ $\lambda 1640$ emission quickly fades and H$\beta$ slowly decays until about 1,148 Myr when all three study volumes simultaneously begin to emit both \heii\ $\lambda 4686$ and \heii\ $\lambda 1640$ lines as well as experience a temporary bump in H$\beta$ emission. After 1,150 Myr, \heii\ emission ceases and H$\beta$ continues to gradually decay except in V1, which is influenced by a second starburst wherein the combined contributions produce stronger \heii\ $\lambda 1640$ and H$\beta$ lines than during the 1,144.6 Myr bump. 

\begin{figure}
\begin{center}
\includegraphics[width=\linewidth]{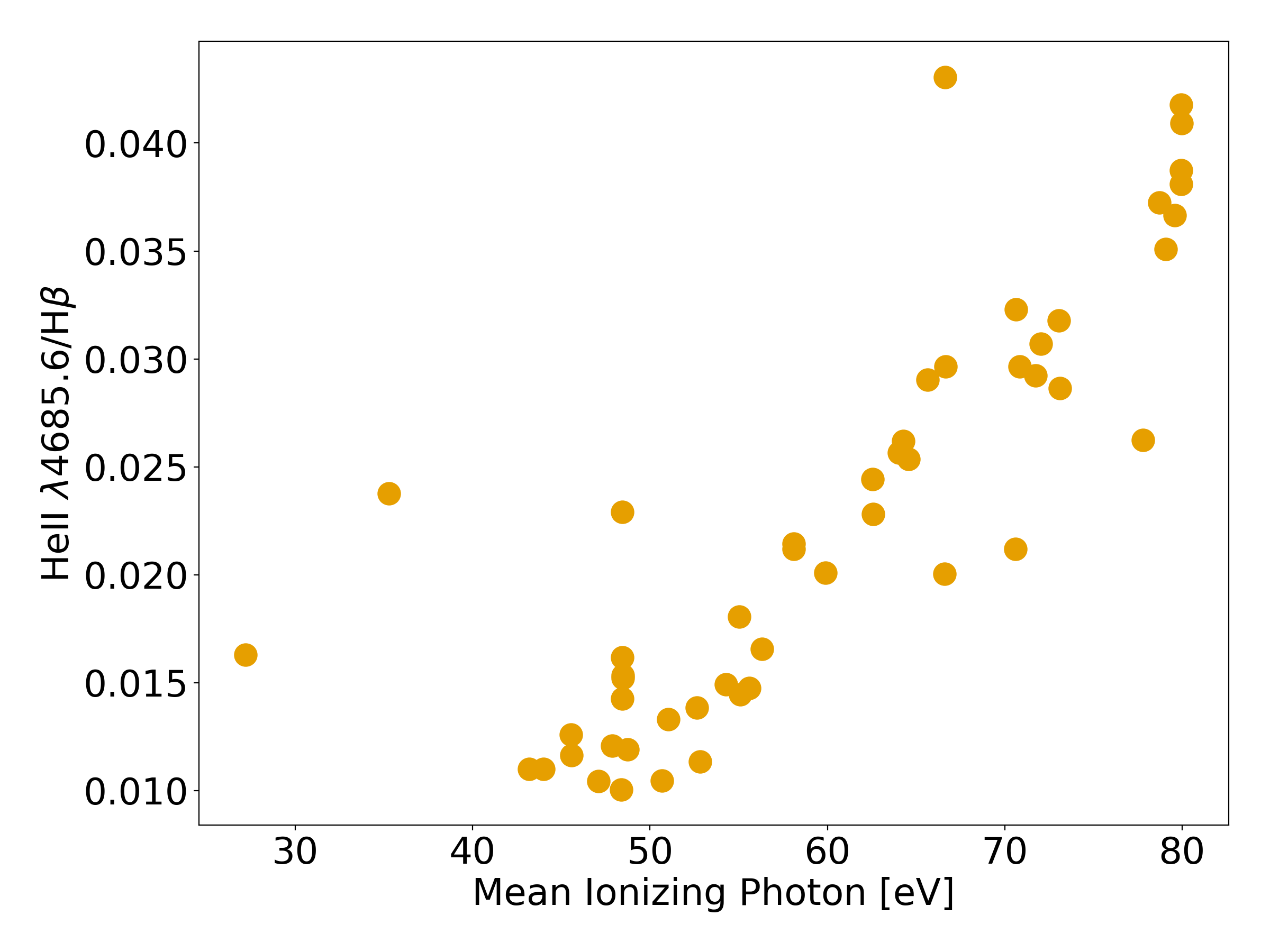}
\caption{Cells with a \heii \ $\lambda 4686$/H$\beta$ ratio greater than 1\% in the 10q 368 kyr timestep simulation versus mean ionizing election energy in eV, excluding low-density outliers.}
\label{fig:meanev}
\end{center}
\end{figure}

\begin{figure}
\begin{center}
\includegraphics[width=\linewidth]{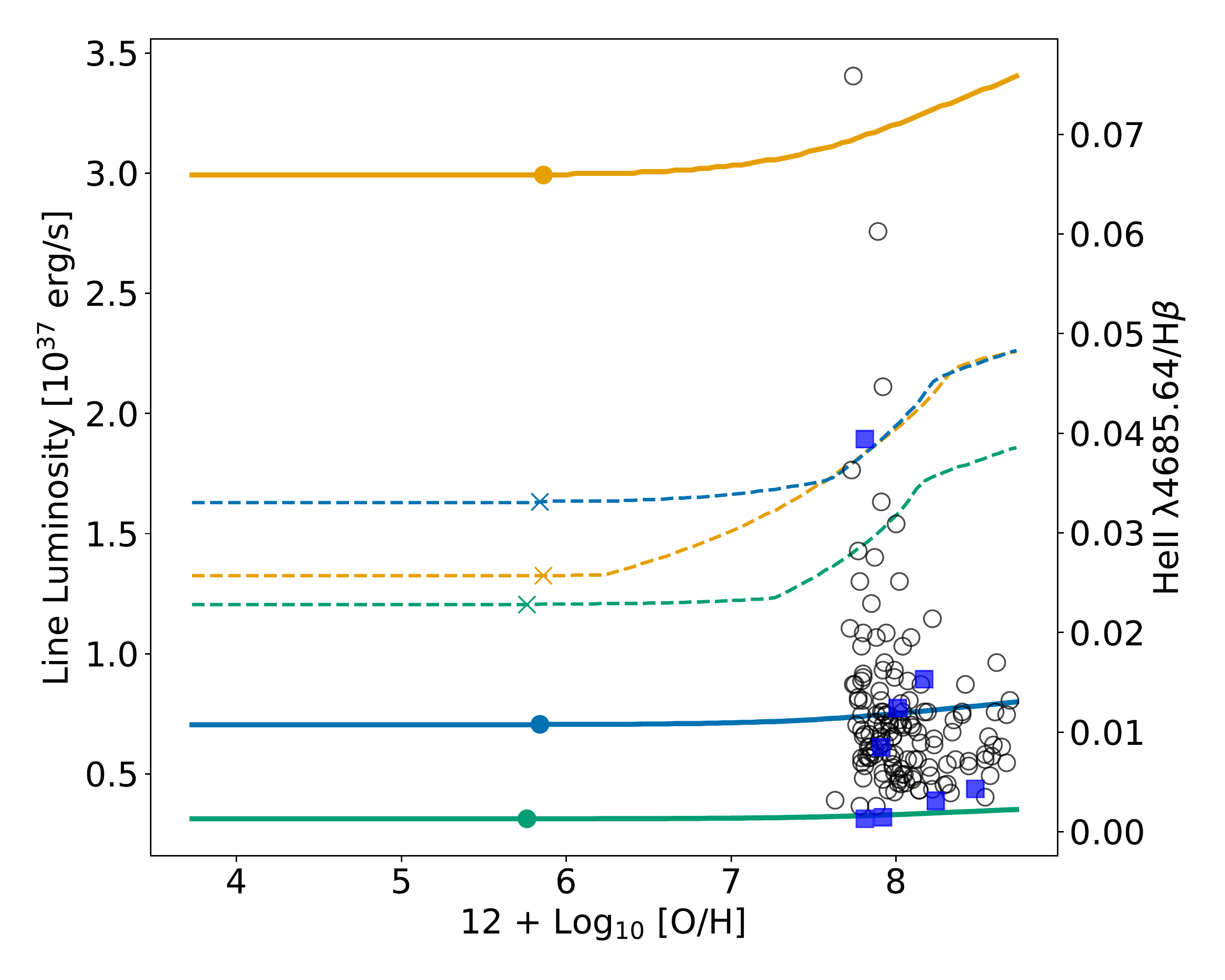}
\caption{Line emission (solid lines, left vertical axis) and the \heii \ $\lambda 4686$/H$\beta$ ratio (right vertical axis, dashed lines) plotted against metallicity. The spectra and density of the gas in the cells during the peak of \heii \ $\lambda 4686$ emission depicted in Figure \ref{fig:threecells} are fixed and the metallicity is changed to explore what the HeHRs would be in a different gas composition. Yellow, blue and green lines represent study volumes V1, V2, and V3. Plot markers along lines indicate the values from the simulation. Black circles are observed galaxy \heii \ $\lambda 4686$/H$\beta$ ratios from \citet{2012MNRAS.421.1043S} with subsolar metallicities. Blue squares are the sample from \citet{2017MNRAS.472.2608S}. Both sets of observations are plotted using their \heii \ $\lambda 4686$/H$\beta$ ratio, which is read off the right vertical axis.}
\label{fig:metals}
\end{center}
\end{figure}

This pattern can be partially explained by the spectral evolution of the stellar population synthesis model. The top left, center left, and bottom left plots of Figure \ref{fig:threecells} show the evolution of the flux as a function of wavelength within V1, V2, and V3 respectively with darker shades depicting the spectra at later times using the same color map as the complimentary time-domain plots in right column of the same row.  H$\beta$  and the \heii\  emission lines is most sensitive to the ionizing continuum and specifically the energy of the mean ionizing photon rather than the total value of $U$. Where the UV intensity of the spectra grows or falls gradually, the ionizing continuum first softens then suddenly hardens midway through the interval due to the presence of WR stars in the FSPS spectra, despite a contemporaneous drop in the ionization parameter. In all study volumes, mean energy of ionizing photons slightly decreases from about 21 eV at 1,144.6 Myr to 17.6 eV until 1,148 Myr, when it suddenly climes to 60 eV, peaks at 80 eV at 1,149 Myr, and then returns back to 18 eV by 1,150 Myr. Given that the ionizing potential of He is 24.6 eV, spikes in \heii\ emission are well explained by the hardness of the spectra. It should be noted that not all of the hardness comes from the population synthesis model. All study volumes are illuminated by the same stellar population and are in close physical proximity, but do not have the same spectral shape or ionization parameter due to the calculation of wavelength-dependent absorption as well as the composition of each cell. For example, in V3, the strength of the soft and hard UV spectrum increases as its density decreases. As such, HeHRs somewhat vary between the volumes. 

The positive correlation between the HeHR and mean energy of ionizing photons is general as shown in Figure \ref{fig:meanev}. All the cells with HeHRs greater than 1\%, have mean ionizing photon energies greater than 25 eV, and more typically greater than 43 eV. Excluded from the plot are outliers wherein the HeHR might exceed 25\% due to extremely low (less than 0.18 nuclei/cm$^3$) gas densities and therefore extraordinarily high ionization parameters (Log $U$ > -1.5). However, these cases do not significantly contribute to the galaxy-wide HeHR because their low density limits the overall luminosity of \heii\ $\lambda 4686$ emission.

Analysis of the 10v simulation during epochs of star formation replicated the trends in the 10q simulation, but also revealed another mode for a HeHRs greater than 1\%. In the largest halo, star formation is quenched between 951 Myr and 1,381 Myr and H$\beta$ emission falls to 50-100 times the luminosity of the \heii\ $\lambda 4686$ background for several tens of millions of years. However, that background should be undetectable in the galactic spectrum (\heii\ $\lambda 4686$ equivalent width less than $10^{-4}\AA$). Importantly,  >1\% HeHRs were produced in the 10v simulation immediately after the first starburst at the end of the several hundred Myr gap in star formation. Since the 10q produced these ratios after 15 Myr gaps in star formation, HeHR ratios are likely not tied to star formation histories older than 15 Myr.

\section{Discussion and Observations}
\label{sec:diss}

\begin{figure}
\begin{center}
\includegraphics[width=\linewidth]{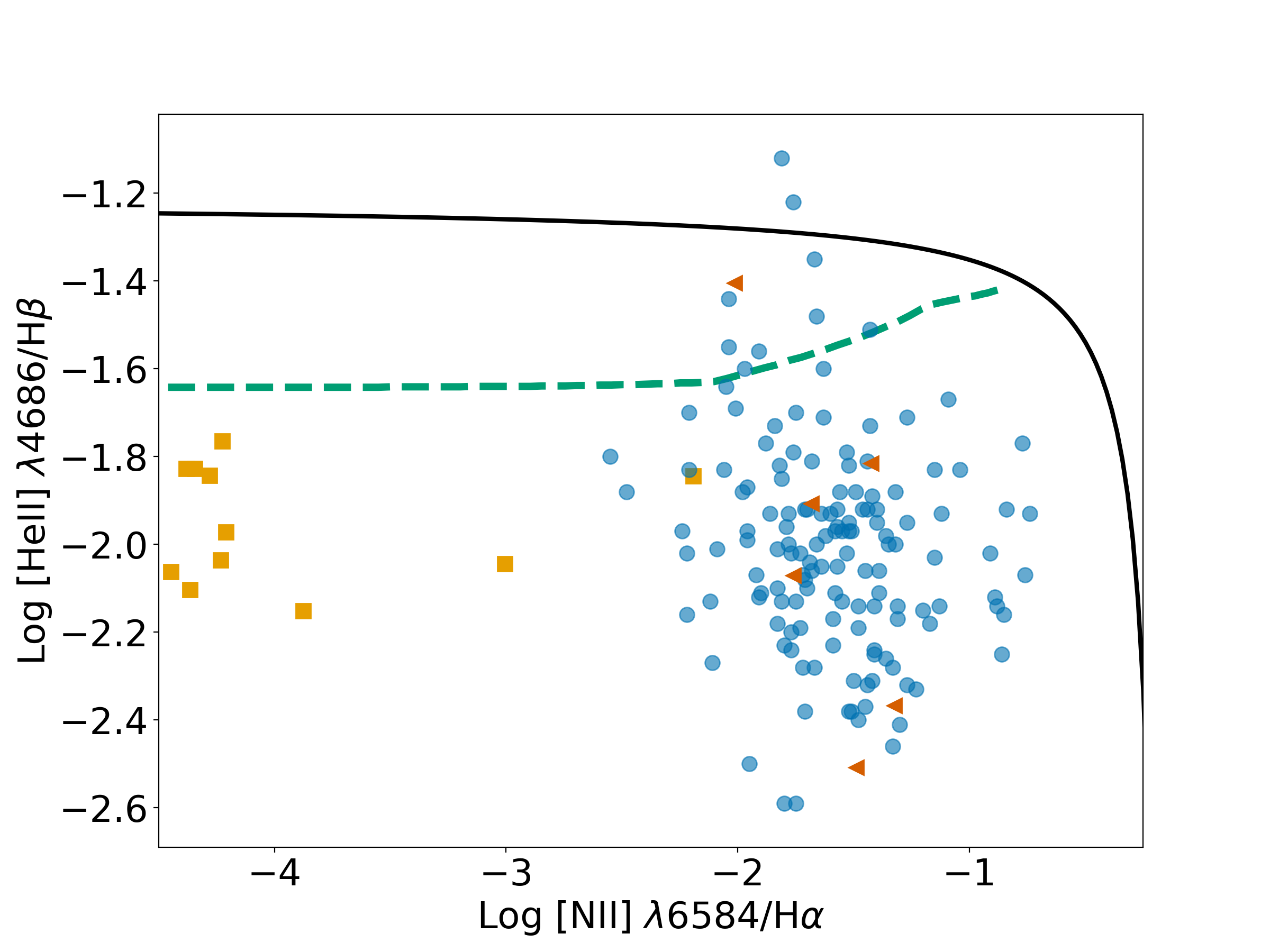}
\caption{\heii \ $\lambda 4686$/H$\beta$ versus \nii\ $\lambda 4584$/H$\alpha$ ratios for peaks in the 10q simulation (yellow squares), the \citet{2012MNRAS.421.1043S} AGN delineation prescription (black line) and sub-solar metallicity galaxy samples (blue circles), and the \citet{2017MNRAS.472.2608S} sample (orange triangles). Points above the line have AGN-like characteristics. The 10q samples are all much lower metallicity, which tends pulls points to the left of the plot. The green dashed line shows the evolution of these ratios in V3 as gas metallicity is increased to solar values, showing the metallicity traversal trend.}
\label{fig:bpt}
\end{center}
\end{figure}

\begin{figure*}
\begin{center}
\includegraphics[width=\textwidth]{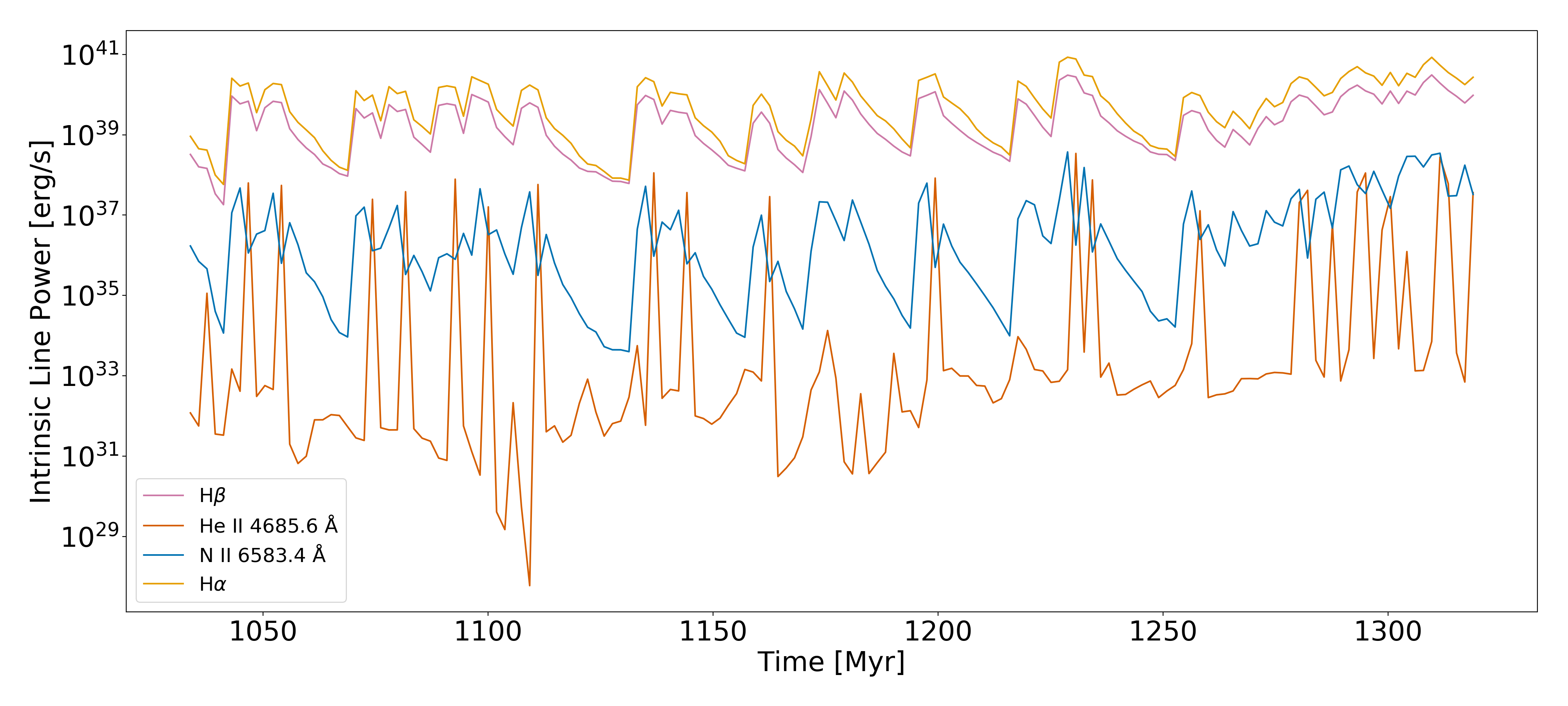}
\caption{Emission lines that comprise the BPT diagram plotted in Figure \ref{fig:bpt} plotted as a function of time. \nii\ $\lambda 4584$ appears to dip during periods of high \heii \ $\lambda 4686$ emission during the low-metallicity, bursty star formation phase and is more luminous during the higher-metallicity, steady star formation phase.}
\label{fig:BPTtime}
\end{center}
\end{figure*}

This modeling technique suggests that galaxy-wide HeHR peaks greater than 1.7\% occur naturally when a few conditions are met. First, \hii\ regions need to be old enough and isolated enough so that a single population of stars with a similar age is the primary source of ionizing radiation. Strong H$\beta$ emission line production around a young star occurs for more than 10 Myr, and \heii\ $\lambda 4686$ production occurs about 5 Myr after the stars form due to the presence of WR stars and persists for a further 1.5 Myr. Therefore, star cluster formation needs to be less frequent than about one in every 15 Myr. This condition seems to be more important than the sSFR, which was not correlated to higher HeHR peaks.  

Secondly, it is more likely to occur at low metallicity, but in a somewhat counter-intuitive manner. To explore how metallicity impacts the HeHR, a range of metallicities from $10^{-5}\ \rm{Z_\odot}$ to 1 $\rm{Z_\odot}$ are used with a fixed density, spectra, and flux distribution taken from the three study volumes at the 1,149 Myr \heii\ $\lambda 4686$ peak to calculate hypothetical photoionization balances and their resulting emission line features. Results are plotted in Figure \ref{fig:metals}. With everything else held equal, the HeHR increases with increases in gas metallicity due mostly to a decrease in H$\beta$ emission with increasing metallicity. However, as shown in Figure \ref{fig:hbratios}, the maximum height of HeHR peaks are generally higher at lower gas metallicities in the 10q simulation. Because high metallicity is a proxy for an active star formation history, high HeHR peaks are less likely at higher metallicities because \heii\ $\lambda 4686$ emission from starburts needs to be strong enough to overcome a stronger H$\beta$ background. As shown in Figure \ref{fig:threecells}, a second episode of star formation can increase H$\beta$ emission by as much as eight- or nine-fold over the contribution of the first, whereas metallicity only offers a twofold improvement in the HeHR. 

That is not to imply that a higher metallicity precludes bursty star formation entirely. \citet{2018MNRAS.473.3717F} conclude that galaxies of any mass have bursty star formation histories at $z > 1$ and all dwarf galaxies have bursty star formations regardless of redshift because gas free fall times are shorter than the supernovae feedback time scales, allowing for star formation to overrun feedback before being quenched. If a higher metallicity halo has conditions similar to the main halo in 10q, HeHRs up to 3\% at gas metallicities of 12 + Log$_{10}$ [O/H] = 7.5 could possibly be explained by the trend in Figure \ref{fig:metals}. However, higher metallicities are also a loose proxy for higher masses in hydrodynamical simulations \citep{2016MNRAS.456.2140M,2019MNRAS.484.5587T}, so simply scaling the conditions in a lower mass halo does not robustly explore the dynamical effects that might lead to differences in star formation and feedback in a potentially larger mass halo. Nonetheless assuming a comparable starburst, most of the sub-solar metallicity HeHR observations between the combined \citet{2010A&A...516A.104L}(not plotted, maximum HeHR = 2.8\%), \citet{2012MNRAS.421.1043S}, and  \citet{2017MNRAS.472.2608S} sample can be reproduced with the conditions in the halo at 1,149 Myr and a gas metallicity adjustment. This remarkably includes HeHR from galaxies that have an AGN-like line ratios or exhibit clear WR features. Figure \ref{fig:bpt} demonstrates how modeled HeHR bursts compare to higher metallicity observations in an HeHR versus \nii\ $\lambda 4584$/H$\alpha$ diagram. Simulation values (yellow squares) generally occupy a space to the left of observations, however this is once again fully explained by the track V3 follows with increases in metallicity (dashed green line), which transverses directly between simulation values and observations. Figure \ref{fig:bpt} also shows that the observations with extraordinarily large HeHRs tend towards the AGN side of the BPT diagram, which might explain their higher ratios. Figure \ref{fig:BPTtime} shows that \nii\ $\lambda 4584$ emission is further weakened during bursty star formation, exhibiting luminosity dips during HeHR peaks as well as during H$\alpha$ dips. Conversely, during the steady star formation phase at the end of the interval, a simulated observation directly overlaps with the observed set despite relatively low simulation gas metallicities. Thus, both metallicity and star formation consistency affect the scatter in the HeHR versus \nii\ $\lambda 4584$/H$\alpha$ diagram.

The third contributing factor is the hardness of the impinging spectrum. A background \heii\ $\lambda 4686$ emission is present in every snapshot of every simulation due the non-zero escape fraction of \heii\ ionizing photons around even old star clusters.  The neutral, low star formation state observed in the 10v simulation supports HeHR ratios greater than 1.3\% but both \heii\ $\lambda 4686$ and H$\beta$ have extremely low equivalent widths. Thus, only the hardness of the spectrum inside denser gas in \hii\ regions contribute to a detectable HeHR peak. Figure \ref{fig:meanev} demonstrates that within \hii\ regions, the small-scale HeHR is positively correlated to the energy of the mean ionizing photon and only becomes relevant once that mean exceeds 25 eV. However, the galaxy-wide HeHR is more sensitive to the overall size and composition of the \hii\ regions and the star formation history. Additionally, the mean ionizing spectrum impinging gas outside of \hii\ regions is not correlated to a strong small-scale or galaxy-scale HeHR.

\subsection{Observational Predictions}

Since the timing of starbursts is related to the magnitude of HeHR peaks, observations and models for high redshift star formation are necessary for making predictions for occurrence rates.  Despite some disagreement about the nature of a sSFRs plateau beyond $z = 2$, observations and models converge \citep{2015A&A...577A.112L} on a range of mean sSFRs between as low as 1.66 Gyr$^{-1}$ \citep{2005ApJ...633L...9F} and as high as 6 Gyr$^{-1}$ \citep{2013ApJ...763..129S} at $4 <z < 5$ for $10^{10}$  $\rm{M_\star}$/$\rm{M_\odot}$ galaxies, with significant scatter between individual sources. This compares to an average sSFR of about 4.18 Gyr$^{-1}$ during the busty period of start formation in the main halo in the 10q simulation between 1034 and 1,240 Myr and 6.9 Gyr$^{-1}$ after 1,275 Myr as plotted in Figure \ref{fig:hbratios}, which has $\rm{M_\star} = 10^{7.48}$ $\rm{M_\odot}$ at $z \sim 5$.

There is some observational evidence that sSFR is independent of $\rm{M_\star}$ at $z =1$ \citep{2012ApJ...754L..14S} and a plot of log(SFR) to log($\rm{M_\star}$) has a slope equal to or higher than $0.65$ at $1.4 < z < 2.6$ \citep{2015ApJ...815...98S}  for $10^{9.5} < \rm{M_\star} < 10^{11}$ $\rm{M_\odot}$. It is somewhat precarious to extend these relationships to higher redshift and lower masses, but if the \citet{2015ApJ...815...98S} SFR-$\rm{M_\star}$ and observed scatter holds to the regime of the 10q simulation, 4.18 Gyr$^{-1}$ for a $\rm{M_\star} = 10^{7.48}$ $\rm{M_\odot}$ halo is 1$\sigma$ below the mean. In the simulation, the HeHR is above 1.3\% for about 4.3\% of the interval during six different star formation events, assuming the HeHR plateaus near their peak for at least $\sim 1.5$ Myr. Under these assumptions, at least about seven out of one thousand galaxies like the main halo in 10q could exhibit a galaxy-wide HeHR above 1.3\% if observed at a random long-term and instantaneous sSFR. Additionally, the main halo of the 10v simulation had a long period without star formation before entering a period bursty formation with at least one HeHR peak greater than 1.3\%, which implies by replication that the 10q main halo is not itself simulating an exceedingly rare process. That is to say that a discovery of a handful of galaxies with an HeHR greater than 1.3\% would not, according to this modeling procedure, lend strong evidence to an exotic or metal-free process. This rate also explains the non-occurrence of an HeHR over 1\% in the Void and Normal regions, which contain a limited number of galaxies of appreciable mass. 

 Observations of HeHR ratios over 1.7\% are implied to be more rare since only two incidents are captured in the interval, but an observed occurrence similarly does not require a special condition other than a temporally-coherent burst of intermediate-metallicity star formation. Importantly, there does not seem to be a condition that the 10 Myr averaged sSFR be large to produce a high or detectable HeHR ratio. In fact, too high of a sSFR would imply the formation of multiple star clusters in proximity and a risk of a lower HeHR due to the timing of the galaxy-wide decay of H$\beta$ emission luminosity. This sSFR-HeHR relationship will be further explored in the context of this emission line modeling technique in the future.

\subsection{Limitations}
 
 HeHR ratios above 3\% and HeHR ratios in galaxies larger than $\rm{M_\star/M_\odot} = 10^{7.48} $, at lower redshifts, or at different metallicities are beyond the explicit scope of the simulations analyzed and will be much more rigorously explored in future investigations along with other emission line trends and ratios. Several more simulations of larger galaxies and to lower redshifts need to be generated before this model can act as a proper aid to observers. This would also require careful tuning and calibration of the hydrodynamical simulation's star formation and feedback prescriptions. Additionally, emission lines calculated in this work have small equivalent widths and might be difficult to directly observe with current facilities. Further simulations to lower redshifts are forthcoming to better overlap with observations in the literature and will be presented in future studies. 
 
This emission line modeling procedure can be improved in various ways. Optical depth calculations within cells would be enhanced by a systematic model rather than a fixed mean path length. Smaller cell sizes for each photoionization calculation would allow for more accurate flux matching from the 5\% level down to the 1\% level. Dust attenuation modeling can be adjusted to better reflect high-redshift dust compositions. Metals can also be broken down by element and tracked as separate phases hydrodynamically. However, these adjustments would likely yield smaller accuracy gains than improvements implemented in this work.

\section{Conclusions}
\label{sec:con}

This work describes a new modeling framework for the photon density distribution used in photoionization calculations. For the first time in the literature, this framework accounts for arbitrary flux distributions that are neatly matched to results from detailed raytracing and attenuation of light from each star in a simulation rather than a model that assumes constant intensity or a spherically symmetric flux distribution around individual stellar sources. This improves the state-of-art of photon density modeling from errors in the order of magnitude down to errors off less than 5\% while simultaneously accounting for photons at 8,000 wavelengths compared to just a few in some other methods. Additionally, ionization parameters are systematically calculated rather than commanded, resulting in a natural determination of ionizing flux throughout multiple, co-contributing and inhomogeneous \hii\ regions.  

Most of this study focuses on using this method to post-process a cosmological radiative-hydrodynamic AMR simulation with minimum cell sizes smaller than 10 pc, metal cooling, radiative feedback, and star formation prescriptions that follow the collapse of molecular clouds to densities over 500 particles/cm$^3$. This level of fidelity allows for a robust determination of star formation histories as well as the dispersion of star clusters represented by multiple star particles.

Though this approach naturally produces a large number of emission line diagnostics, this analysis focuses on galactic \heii \ $\lambda 4686$/H$\beta$ emission line ratios (HeHRs) as a test of the method, since that ratio has been the subject of much study and debate. The following trends were uncovered by following the dynamical evolution of $\sim \rm{10^{10}\ M_\odot}$ galaxies at redshifts $4 < z < 5.5$:

\begin{itemize}

\item Episodes of bursty star formation that occur naturally in the simulation produce star particles with formation times close enough to promote a coherent burst of \heii \ $\lambda 4686$ emission, similar to but not equivalent to an instantaneous burst model in a stellar population synthesis model.

\item The strength of the HeHR following a starburst is modulated by the strength of H$\beta$ emission and the number of  \hii\ regions. H$\beta$ emission in an \hii\ region decays with time and about 15 Myr between starbursts is needed for \heii \ $\lambda 4686$ emission to exceed 1\% of the H$\beta$ background. 

\item Since bursty star formation is a feature of dwarf galaxies and high-redshift galaxies, predictions for observed population of \heii \ $\lambda 4686$ emitters are principally modulated by the short length of strong \heii \ $\lambda 4686$ emission after each starburst. HeHRs of as much as 1.7\%, though relatively rare, are well explained by processes naturally seen in simulations. New observations of high redshift galaxies are likely to have HeHR peaks in the 1-2\% range without requiring exotic processes.

\item Though higher gas metallicities can produce higher HeHRs in individual calculations, the effect is less important than maintaining the proper interval between starbursts and the strength of the H$\beta$ background. Low metallicity halos are seen producing HeHRs that exceed most of the observed sub-solar sample. Therefore, the downward slope in HeHR for metallicities 12 + Log$_{10}$[O/H] > 7.75 may be a question of a high-metallicity suppressing mechanism rather than a low-metallicity enhancing mechanism. More low metallicity observations of HeHRs over 3\% would challenge this notion in the absence of a corresponding simulated observation. 

\item Volumes of gas with extraordinarily high ionization parameters (Log $U > -2.3$) did not strongly contribute to modeled galaxy-wide HeHRs because their densities limited the total \heii \ $\lambda 4686$ luminosity from those regions. The mean energy of ionizing photons was much more strongly correlated ($r=0.837$) to the HeHR ratio than ionizing parameter ($r=-0.0781$).

\item The conditions for the strong HeHRs in these simulations are likely not properly or completely captured by single source, fixed density, or spherical flux assumptions in other models.

\end{itemize}

This work establishes an emission line modeling framework for the {\sc Caius} pipeline. Future studies will extend to more regimes and epochs with goal of building a truly generalized radiative transfer platform for synthetic observations.

\section*{ACKNOWLEDGMENTS}

This work was supported by XSEDE computing grants TG-AST190001 and TG-AST180052 and the Stampede2 supercomputer at the Texas Advanced Computing Center. KSSB was supported by a Porat Postdoctoral Fellowship at Stanford University.

\bibliography{ResearchStatement}
\bibliographystyle{aasjournal}

\bsp
\label{lastpage}

\end{document}